\documentclass[11pt]{article}
\usepackage{amsthm}
\usepackage{url}
\usepackage{amssymb}
\usepackage{amsfonts}
\usepackage[english]{babel}

\usepackage{geometry} 
\usepackage{color}
\usepackage{graphicx}
\usepackage[capbesideposition=outside,capbesidesep=quad]{floatrow}
\usepackage{url}
\begin{document}
\title{Large-scale Species Tree Estimation}
\author{Erin Molloy and Tandy Warnow}
\maketitle{}

\abstract{Species tree estimation is a complex problem, due to the fact
that different parts of the genome can have different evolutionary
histories than the genome itself. 
One of the causes for this discord is {\em incomplete lineage sorting}
(also called deep coalescence), which is a population-level process
that produces gene trees that differ from the species tree.
 The last decade has seen a large number of
 new methods developed to estimate species trees from
 multi-locus datasets, specifically addressing this cause of
 gene tree heterogeneity.
 In this paper, we review these methods, focusing
 mainly on issues that relate to  analyses of datasets 
 containing large numbers of species or 
 loci (or both).
We also discuss  divide-and-conquer strategies  for enabling species tree estimation methods to  run on  large datasets, 
including new approaches that are based on algorithms (such as TreeMerge) for the ``Disjoint Tree Merger"  
problem. 
 }
\section{Introduction}
Phylogenetic trees, whether for genes or species, form a basis by which many biological questions can be
addressed.
Species trees in particular are used in many biological studies  to address
biological questions of interest, including the evolution of traits. 
The accuracy of the estimated species tree can therefore have an impact on
the accuracy of the biological research using that tree.

In practice,  phylogenies (whether species trees or gene trees) are nearly always estimated using statistical methods, such as 
maximum likelihood,  that take as  input  a multiple sequence
alignment and assume that all the sites evolve down some common (but
unknown) tree under some sequence evolution model, such as the Generalised Time Reversible (GTR) 
model \cite{tavare-gtr}.
These approaches can be statistically consistent when the alignment is correct
and the true evolutionary model fits the assumed evolutionary model. 
For example, maximum likelihood, if solved exactly, is statistically consistent
under the GTR model  but maximum parsimony is not~\cite{Felsenstein1978}.
However, not all such conditions guarantee statistical consistency,
since the {\em no common mechanism}   model of Tuffley and Steel is not
identifiable \cite{tuffley.steel}, and so even maximum likelihood under that model is
not statistically consistent.
Hence, statistical consistency for the phylogeny estimation depends, among 
other things, on the
evolutionary process being sufficiently well-behaved. 

These assumptions---that the alignment is correct, that all the sites in the input alignment evolve down a single 
tree, and that the statistical model governing the evolution of the sites is sufficiently  well behaved---are
significant, and may not hold for any given dataset.
For example, multiple sequence alignment can be very difficult, especially on 
large datasets \cite{sate2009,Mirarab2014,NguyenUPP2015,Morrison-MSA-unsolved}, 
and  biological evolution is much more complicated than standard sequence evolution models (e.g., these models do not include heterotachy, selection,  dependencies between sites, etc.~\cite{Philippe-Heterotachy,Pupko-models,kubatko-mpe2016}).

Another basic challenge is that many biological processes, such 
as gene duplication and loss \cite{nadia-festschrift-chapter}, incomplete lineage sorting \cite{maddison1997}, horizontal gene transfer \cite{nakhleh-festschrift-chapter},
and hybrid speciation \cite{hyde2018}, 
cause different parts of the genome to have different histories, with the consequence that
the true evolutionary tree  for any single genomic region may not match
the true species tree!
As a result, species trees are usually estimated by using information from different
genes, in the hope that the information across all the genes will help identify the species tree.
In other words, {\em multi-locus} species tree estimation is used rather than single locus approaches.
Sometimes these multi-locus datasets encompass much of the genomic data, and
so can even be considered genome-scale datasets.

Although there are multiple sources for discord between gene trees
and species trees, much of the focus in terms of method development
has been for species tree estimation in the presence of
incomplete lineage sorting (ILS), as modelled by the multi-species coalescent
\cite{kingman}.
Under this model, the true species tree is a rooted binary tree with leaves labelled
by species, the internal nodes represent speciation events, 
and branch lengths are given in coalescent units.
Each model species tree defines a distribution on gene tree topologies,
and every possible gene tree topology has strictly positive probability of
being generated by the species tree.
Furthermore, the species tree topology and branch lengths are identifiable from
the distribution on gene trees~\cite{ADR2011a}. 

Many different methods have been developed to estimate
species trees from multi-locus datasets~\cite{Edwards-new2009,knowles2009estimating,kubatko-review2009,DegnanRosenberg2009,Molloy2018}.
Broadly speaking,  there are three different types of species tree estimation
methods that address gene tree heterogeneity resulting from ILS.
The first type operates by computing gene trees (one tree for every genomic
locus) and then uses the set of gene trees to compute a species tree.
Since these methods operate by using summary statistics to estimate the species tree,
they  are
 called ``summary methods".
 The most well-known such method is ASTRAL \cite{astral,astral2,astral3}, 
 but other summary methods include NJst \cite{njst}, MP-EST \cite{mpest}, and ASTRID \cite{astrid}.
The second type co-estimates the gene trees and species tree, with
*BEAST \cite{starbeast} and STARBEAST-2 \cite{starbeast2} the most well-known examples.
The third type uses the
site patterns in the concatenated alignment, and so bypasses entirely the
challenge of estimating gene trees; SVDquartets \cite{svdquartets} is the most
well-known example of these ``site-based methods".
Although these methods use different techniques, they have been proven statistically consistent under
the multi-species coalescent (MSC) model \cite{kingman}.

Despite the availability of methods for species tree estimation that explicitly
address gene tree heterogeneity, 
the most common approach simply concatenates
the different
alignments for the different loci into one ``concatenated alignment" (also called
a ``super-alignment") and
then constructs a tree on that larger alignment
using standard phylogeny estimation methods, such as 
maximum likelihood.
However, 
maximum likelihood on the concatenated alignment can be statistically
inconsistent (and even positively misleading) when there is
gene tree discord as a result of ILS, as
established in \cite{RochSteel} (for unpartitioned maximum likelihood)  and
\cite{RochNuteWarnow2018}  (for fully partitioned maximum likelihood).

Studies comparing these concatenation analyses (often computed using
heuristics for maximum likelihood, such as
RAxML \cite{Stamatakis2006,raxml8}) to the 
coalescent-based species tree methods described above have
shown mixed performance.
As would be expected, when ILS is low enough, the better
concatenation analyses tend to produce the most
accurate trees and when ILS is high enough
then coalescent-based methods can be more accurate \cite{Molloy2018}.
The performance of concatenation
when ILS is very high, and in particular in the presence of the  {\em anomaly zone} (i.e.,
where the most probable gene tree is different from the species tree \cite{degnan2006discordance,degnan2013,Rosenberg2013-agt}), has been a specific concern \cite{kubatko-degnan-2007}. 
Although concatenation  often has poor accuracy in the anomaly zone,
even here there are even some  cases (again, typically characterized by genes with very low
phylogenetic signal)
where concatenation analyses are  more accurate  than the best summary methods~\cite{Molloy2018}.
Thus, from an empirical viewpoint, the choice of method is complex, and depends on
the phylogenetic signal in the input as well as possibly other properties of the data.

From the perspective of theory, the issue is also complex, as
most species tree estimation methods become positively misleading when the number of 
sites per locus is bounded, even if the number of loci
is allowed to increase without bound~\cite{RochNuteWarnow2018}.
This cautionary statement has been proven for fully partitioned maximum likelihood and
standard summary methods, and may also apply to co-estimation methods!
Thus, the choice of whether to use a coalescent-based approach
or concatenation analysis is by no means straightforward.

Having said this, many of the current generation of
species tree estimation methods  are  
highly accurate, and in some conditions they are more accurate
than traditional methods based on concatenation \cite{Molloy2018}.
Hence, at a minimum, they provide an alternative technique for
estimating
 the species tree that can be used in concert with other approaches
 (such as traditional concatenation analysis)
 to identify those regions within the evolutionary tree
that are supported by multiple types of phylogenetic analysis.
Importantly, as we will show, many of these ILS-aware
species tree estimation methods {\em have computational
advantages over concatenation analyses}, especially when
the concatenation analysis is based on maximum likelihood or
Bayesian MCMC.
Thus, ILS-aware species tree estimation methods 
can provide a surprisingly fast and often highly accurate
estimate of species trees, even on very large multi-locus datasets. 
This chapter explores the computational aspects of the
leading current species tree estimation methods,
with an eye towards the future development of new methods.

The rest of the chapter is organized as follows.
We begin in Section \ref{sec:species-tree-methods} with a discussion of selected 
methods that are used for multi-locus species tree estimation, focusing on their
computational performance. 
We describe divide-and-conquer approaches that have been
used to scale these methods to larger datasets in Section \ref{sec:divide-and-conquer}.
The impact of parallel computing on scalability is discussed in Section \ref{sec:parallel}. 
We compare  the current
 methods in Section \ref{sec:discussion} with respect to their relative performance and utility for species
 tree estimation on large datasets.
 We conclude in Section \ref{sec:conclusions} with a summary of
 our observations as well as a discussion of
opportunities for future research. 
Finally, Appendix A includes a discussion of ``Big-O" notation and how it is used
in running time analyses, and Appendix B provides a list of the basic information about the software packages
implementing these methods, including links to 
webpages  to obtain the software.

\section{Species tree estimation methods}
\label{sec:species-tree-methods}

We now discuss specific methods for each of the major types of species
tree estimation methods in use today, focusing on scalability and accuracy
where  the number of loci and/or number of species is large.
We mainly explore  two maximum likelihood heuristcs
(RAxML \cite{Stamatakis2006} and FastTree \cite{Price2010}),
two summary methods (ASTRAL-III \cite{astral3} and ASTRID \cite{astrid}), and
two site-based methods (SVDquartets \cite{svdquartets}, as implemented in
PAUP* \cite{paupstar}, and SVDquest \cite{svdquest}).
We also  discuss  a co-estimation method (*BEAST \cite{starbeast}), but as
its computational requirements limit it to small data sets, our discussion is fairly
limited.
With the exception of ASTRID and SVDquest, these are  methods
that are well known in the systematics community.
We also discuss techniques  that have been
designed for improved scalability:
BBCA \cite{bbca}, which is a modification of *BEAST to enable it
to run on large numbers of loci, and
NJMerge \cite{njmerge,njmerge-biorxiv}, which is a generic technique for
scaling species trees methods to large numbers of species.

For all the methods we discuss, the input is a multi-locus dataset where each
locus is given either as a multiple sequence alignment or as a tree computed from the
alignment for the locus. 
Note also that we do not constrain the loci to all have the same set of species.
Furthermore, for the purpose of this chapter, we will assume that each locus has at most
one copy of any species, though not all methods have this as a requirement. 
Finally, the output tree will have one leaf for every species that appears in any input locus.

\subsection{Concatenation using maximum likelihood}
Maximum likelihood (ML), which seeks the model tree (under the specified stochastic model), is
a standard approach in phylogeny estimation, but is NP-hard \cite{rochnphard}.
Hence, methods for ML, such as RAxML, IQ-TREE \cite{iqtree}, and PhyML \cite{Guindon2003}, use
local search strategies to find good solutions to the maximum likelihood optimization problem.
ML heuristics typically operate by perturbing a tree (e.g., using NNI or SBR moves) and then, depending on its likelihood score, searching from either the original or the perturbed tree.
This process repeats until a stopping criterion is met, typically meaning that the tree has converged to a local optimum.
Because the running time of this approach is non-deterministic, a Big-O running time analysis cannot be provided.
However, a large number of search moves may be required to reach convergence, especially for datasets with large numbers of species, and even the likelihood calculation can be computationally intensive for large alignments, especially if model parameters (e.g., branch lengths) must also be optimized.

In practice, ML heuristics tend to have large memory and running time requirements, but they can produce species trees with high accuracy even when there is gene tree heterogeneity due to ILS \cite{Molloy2018}.
However, from a practical viewpoint, ML heuristics are among the most computationally 
intensive of the approaches we consider, at least on large datasets (large numbers of loci
and/or large numbers of species).
For example, the ML concatenation analysis of the Avian Phylogenomics Project
dataset, which had 48 species and about 14,000 loci (and many millions of sites),
took more than 200 CPU years and at least 1Tb of memory \cite{jarvis-2014b}.

FastTree is an outlier among ML heuristics. Although it is also a heuristic for maximum likelihood (in that it is not guaranteed to find an optimal solution), 
it explicitly limits its search to guarantee that it ends after a polynomial number of
NNI moves. As a result, it is much faster than the other heuristics.
In fact, it is fast enough to run on datasets with hundreds of thousands of sequences.
FastTree was used to construct a tree on alignments computed by PASTA  \cite{Mirarab2014} and UPP \cite{NguyenUPP2015} on the million-taxon RNASim dataset. 
To our knowledge, there is no comparably fast ML heuristic that can analyze
ultra-large datasets.

A comparison between RAxML and FastTree is quite
interesting. 
FastTree is much faster than RAxML on large datasets, for the obvious 
reason: FastTree explicitly limits the number of NNI moves it performs, whereas RAxML does a much better search for ML scores.
This difference means that RAxML will take longer than FastTree, but the result is
that RAxML consistently produces trees with better ML scores than FastTree.
Whether these differences in ML scores translate to meaningful differences in tree accuracy is unclear.
In the simulation study reported in  \cite{liu-fasttree}, 
the trees produced by
FastTree and RAxML were approximately the same in terms of accuracy
and topological differences, when they appeared, were minor.
Thus, there may be  conditions where it is (reasonably) safe to use
FastTree as a technique to compute a large phylogeny (meaning it 
will produce a tree that has reasonable topological accuracy, comparable to what
RAxML or some other heuristic would find), but
it is clear that there are other conditions where the trees it finds
may be substantially less accurate than those found by better
heuristics.
More research is needed to explore this issue.

\paragraph{Summary. }
Concatenation analyses using maximum likelihood remain the
most common approach to estimating species trees
from multi-locus datasets.
They can be quite accurate even in the presence of heterogeneity between gene trees and species trees due to ILS, and sometimes can be more accurate than the ILS-aware methods we discuss here.
However, from a computational perspective,
the better heuristics for ML are very expensive---prohibitively so when the number of loci and number of
species are both large.
Indeed,  the cost of analyzing some multi-locus datasets with fewer than 100 species, such as the Avian phylogenomics dataset \cite{jarvis-2014b},
can be  prohibitive without access to a
supercomputer. 
We discuss the computational challenges in using concatenation analyses using 
supercomputers
later in Section \ref{sec:parallel}.

\subsection{Summary Methods}
\label{sec:summaryl}

Summary methods operate by combining gene trees together and are typically 
very fast, once the gene trees are computed.
Because of their speed, they are a very popular approach for species tree
estimation. 
In this section, we discuss two summary methods: 
ASTRAL, introduced in \cite{astral}, and ASTRID \cite{astrid}.
Both have very good accuracy and speed, and yet use very 
different algorithmic approaches. 
Here we describe these methods and their computational aspects, assuming that
gene trees are already computed.

The main thing to realize, with respect to computational requirements, is that when using one of these fast summary methods, the vast majority of the time is typically spent computing gene trees.
As discussed in Section \ref{sec:parallel}, the computational requirements for computing gene trees are typically less than the computational requirements for concatenation analyses.
Thus, for phylogenomic datasets with thousands of loci, summary methods can have a clear computational advantage over concatenation analyses.
Furthermore, in many biological analyses the gene trees are of independent interest, and so they are computed anyway; thus,  summary methods are generally a low cost part of a computational pipeline for phylogenomic species tree estimation.

\subsubsection{ASTRAL}
ASTRAL was introduced in \cite{astral}, and
subsequently modified and improved in~\cite{astral2,astral3}.
It is  now one of the most commonly used methods
for species tree estimation that address ILS.
The  version of ASTRAL on Github on this date (December 31, 2018) is  ASTRAL 5.6.3. 
There are many differences between the various versions of ASTRAL,
which are described in the chapter by Siavash Mirarab, in this volume.
For the purposes of this chapter, we will focus on the aspects of the ASTRAL algorithm that
impact the running time and scalability to large datasets.
We continue with a discussion of the first version of ASTRAL as described
in \cite{astral}.

The input to ASTRAL  is a set $\mathcal{T} = \{T_1, T_2, \ldots, T_k\}$ of gene trees.
Letting $\mathcal{L}(t)$ denote the leafset of tree $t$ and $S = \cup_i \mathcal{L}(T_i)$,
ASTRAL returns a tree $T$ 
satisfying  $\mathcal{L}(T) = S$, with $n=|S|$.
Thus, the input to ASTRAL has size $O(nk)$ (since there are $k$ trees, each with at most
$n$ leaves).


Letting $T$ be a species tree on $S$ and $T_i$ be a gene tree on $S_i \subseteq S$,
the number of quartet trees they share is given by $|Q(T) \cap Q(T_i)|$, where
$Q(t)$ denotes the set of quartet trees in a tree $t$.
Note that each edge $e$ in a tree $t$ defines a bipartition $\pi_e=(A,B)$ on its leaf set 
(i.e., deleting $e$ from $t$ produces two sets of leaves $A$ and $B$).
Let $Bip(T_i)$ denote the set of bipartitions of $T_i$ (i.e., $Bip(T_i) = \{\pi_e: e \in E(T_i)\}$).

When $\mathcal{L}(T_i)=S$, then 
every bipartition of $T_i$ is a ``full bipartition" of $S$ (in that there are no 
missing species).
We express this by saying that $T_i$ is ``complete".
On the other hand, if  one or more species in $S$ are missing from $\mathcal{L}(T_i)$,
then none of the bipartitions of $T_i$ are  bipartitions of $S$.
Any tree $T_i$ that is missing one or more species of $S$ is called ``incomplete".

In the first step, ASTRAL uses $\mathcal{T}$ to compute a set $X$ of ``allowed bipartitions".
The later versions of ASTRAL have varied in how the set $X$ is
computed from the input, but the first version sets
$X$ to contain exactly  the  bipartitions in $Bip(T_i)$ for any $T_i$ that is complete. 
Note therefore that if every tree $T_i$ is complete, then the size of $X$ can 
be as large as $k(n-3)+n$, where $n=|S|$ (since it is possible for all the trees to
differ on all the internal edges, and so only share the trivial bipartitions defined
by leaves).

Now that we have defined all these concepts, we can state what
ASTRAL does. 
Given the input set $\mathcal{T}$ of gene trees, 
ASTRAL computes the set $X$ of allowed bipartitions, and then
solves (exactly!) the following optimization problem:
Find a tree $T$ with leafset  $S$ where
$$T = \arg \max \sum_{i=1}^k |Q(T) \cap Q(T_i)|$$
subject to the constraint that $Bip(T) \subseteq X$.

In other words, ASTRAL finds a species tree $T$ that
agrees with as many of the quartet trees induced by the gene trees in $\mathcal{T}$ as
possible,  but where $T$ is
within a constrained search space defined by $X$.
If $X$ is the set of all possible bipartitions on $S$, then
ASTRAL is seeking the tree that agrees with as many
quartet trees as it can, which is unfortunately an NP-hard problem
\cite{LafondScornavacca2016}.

ASTRAL finds this optimal tree
using dynamic programming (an algorithmic technique).
The analysis of the running time for ASTRAL-1 \cite{astral} was
shown to be $O(|X|^2 n^2k)$, 
and ASTRAL-2 \cite{astral2} improved the running time to $O(|X|^2 nk)$ time.
Later analyses brought the running time down somewhat (by decreasing the
exponent for $|X|$). 
Hence,
the running time for current versions of ASTRAL is linear in the number in the number of gene trees
and in the number of species, but close to quadratic in $|X|$, the size of
the bipartition space.

Note also that if all the trees in $\mathcal{T}$ are identical (i.e., there is some tree $T^*$ so
that  $T_i = T^*$ for
all $i=1,2,\ldots,k$), then
 $X = Bip(T_1) = Bip(T_2) = \ldots = Bip(T_k)$, and so $|X|=n-3$.
 Therefore, by the definition of ASTRAL's optimization problem (which is
 constrained to produce a  tree satisfying $Bip(T) \subseteq X$),   
 ASTRAL will return the unique tree $T^*$ in $\mathcal{T}$ and its running time
 will be very fast.
On the other hand, if all the trees in $\mathcal{T}$ are very different,
then $|X|$ will be much larger (potentially as big as $k(n-3)$, and the number of legal outputs for
ASTRAL will increase and its running time will increase.
 Thus, the more heterogeneity among the gene trees the larger the space of
legal outputs for ASTRAL, and the
more likely
that ASTRAL's running time will increase.

It is worth considering the causes for heterogeneity in the input set $\mathcal{T}$ that result in
$|X|$ being large.
One natural cause is heterogeneity due to ILS, or other biological processes that make true
gene trees different from the true species tree and from each other.
But heterogeneity in $\mathcal{T}$ can also be explained by {\em gene tree estimation error},
which can result from many causes, including inadequate phylogenetic signal (due perhaps
to insufficient sequence length), model misspecification in the gene tree model, errors in the sequence
alignment used to construct the tree, insufficient running time to find a good tree  (perhaps due to terminating the tree search before convergence to a local optimum),  etc. 
Thus, heterogeneity in $\mathcal{T}$ can result from several causes, and substantial
heterogeneity is generally to
be expected {\em except} when ILS is very low and gene tree estimation error is {\em also} very low.
These two conditions can occur in some datasets, but genome-scale datasets are unlikely to
exhibit both properties, at least in part because many loci will be slowly evolving and so have
inadequate signal, or will evolve so quickly that alignment estimation becomes difficult.
In other words, in many (and perhaps most) multi-locus datasets, heterogeneity will be high,
and $|X|$ will be large.

The later versions of ASTRAL expanded the set $X$ to ensure that
a larger part of the possible treespace was explored.
This improved accuracy but also increased the running time.
Subsequently, various algorithmic techniques were developed to reduce the
running time, including techniques that operated by not expanding $X$ quite
as much. The  current version (i.e., ASTRAL 5.6.1) has been well optimized,
and can run on many large datasets.
However, even this current version is impacted by the
heterogeneity in the set of gene trees, since it requires
that $X$ contain all the bipartitions found in all the complete gene trees.
Hence, for some inputs, ASTRAL will be slow. 
For example, there are  conditions with very high ILS, explored in \cite{njst}, in which ASTRAL v.5.6.1.~did not complete on
some datasets with 1000 species and 1000 loci, even given 48 hours.   
However, there are other datasets of the same size, but lower heterogeneity, 
where ASTRAL
completes quickly on (i.e., in a few hours) \cite{astral2,astral3}.
Thus, ASTRAL's running time  significantly depends on the characteristics of the input data.


\paragraph{Summary. }
ASTRAL is one of the best summary methods in terms of accuracy.
In addition, it can scale to very large datasets with hundreds to thousands of species and
loci. 
Although the running time depends on the properties of the dataset, there are many
datasets for which it is quite fast.
It is a good option to consider when the number of loci is very large, as it
can be faster than concatenation analyses under many circumstances and
often provides comparable accuracy (and is sometimes more accurate).

\subsubsection{ASTRID}

ASTRID is a summary method that was introduced in \cite{astrid}.
Like ASTRAL, 
the input to ASTRID is also a set $\mathcal{T}$ of gene trees,
and the output is a species tree that contains  one leaf for
every species in $S$.
Note that the current version of ASTRID 
can only handle inputs where each gene tree has at most one
copy of each species. 

ASTRID is very similar to NJst \cite{njst}, with 
essentially only one change, as described below.
In the first step, ASTRID computes the {\em average internode distance 
matrix}, which is the $n \times n$ matrix $D$ where $D[i,j]$ is the
average of the distances between leaves $i$ and $j$ across all
the trees in $\mathcal{T}$ that contain both $i$ and $j$.
 ASTRID seeks a tree under the balanced 
minimum evolution (BME) criterion using FastME \cite{fastme-2015} if there are no missing entries, and
otherwise ASTRID runs BIONJ* \cite{phyd}.

NJst is identical to ASTRID except that it runs neighbor joining \cite{NJ} on
the internode distance matrix.
Because neighbor joining requires
all the entries of the distance matrix to be non-empty, this restricts
NJst to only those inputs with no missing entries in the internode
distance matrix (note that entry $D[i,j]$ is undefined whenever none of the trees in 
$\mathcal{T}$ contain both $i$ and $j$, and this is what makes the internode
distance matrix have missing entries).

Thus, one main difference between ASTRID and NJst is that ASTRID can run
on all datasets (because BIONJ* is designed for datasets with  missing entries),
but NJst will fail to run on some datasets.
On those datasets where both methods can run, NJst runs neighbor joining, and ASTRID runs BME within FastME; thus, the difference in accuracy depends on whether NJ or BME within FastME is more accurate. 
Although it is not completely clear when one method will be
more accurate than the other, the evidence generally suggests
that BME within FastME typically has an empirical advantage over neighbor joining \cite{DesperGascuel2004,astrid,wang-jme2006}.
Finally, given the input set of gene trees, the running time for ASTRID is $O(kn^2)$ to compute the
internode distance matrix, and then $O(n^2 \log n)$ to run FastME \cite{fastme-2015}. 
Hence, the total time of ASTRID is $O(n^2 (k + \log n))$.

\paragraph{Summary. }
Although there have not been extensive studies evaluating ASTRID, results from
simulation studies reported  in \cite{astrid,Molloy2018,Nute-missing2018} suggest that
ASTRID is competitive with ASTRAL-II in terms of accuracy:
sometimes ASTRID is more accurate and sometimes ASTRAL is
more accurate. 
Unfortunately, there are no published results comparing ASTRID to ASTRAL-III.
Nevertheless, given ASTRID's speed, it is a safe method to include in a phylogenomics analysis.

\subsection{Co-estimation methods}
Another type of multi-locus species tree estimation method co-estimates the
gene trees and species trees under processes that take into account
gene tree heterogeneity due to ILS.
 *BEAST,  perhaps the most
well-known of these co-estimation methods, is a Bayesian method that uses MCMC to
sample from the distribution on gene trees and species trees.
Hence, the input to *BEAST is a set of
gene sequence alignments, and the output is a set of gene tree distributions (one distributions
for each locus)
and a distribution of species trees. 
Then, given these distributions, a point estimate for each gene tree can be obtained,
and a point estimate of the species tree can also be made.
*BEAST  is generally too computationally intensive to use except on 
moderate-sized datasets, and datasets with
100 or more genes and 50 or more species may require several months of CPU time
\cite{mccormack2013,naive-binning,leavitt2016}.
Recently, a new version of  *BEAST has been developed, called StarBEAST2 \cite{starbeast2}, that
 is implemented for greater efficiency.
 As noted in \cite{starbeast2}, StarBEAST2 is reported to be 33 times faster
than *BEAST on simulated datasets. 
Given the tremendous interest in co-estimation of species trees and gene trees,
this reduction in running time is definitely progress towards making co-estimation
feasible on larger datasets.

\paragraph{Summary}
Methods that can co-estimate species trees and gene trees are more computationally
intensive than all the methods discussed here.
Because of the MCMC technique, they are unlikely to scale to large numbers of 
species; the current limit for even StarBEAST2 is likely to be at most 50 species.
The number of loci also impacts the running time, and many analyses have been
based on limiting the analyses to a subset of the loci that is small enough for the
analysis to complete with good ESS values. 
In general, the current co-estimation methods, although highly accurate when
they can be run, are the least scalable of all the methods we discuss here, and
so are not suitable for analyzing genome-scale phylogenomic datasets or
datasets where the number of species is 100 or more.

\subsection{Site-based methods}
The class of methods we consider in this section use statistical properties of the
multi-species coalescent model to estimate the species tree, and they do this
directly  from the concatenated multiple 
sequence alignments.
However, unlike summary methods, site-based methods do not first estimate gene trees, and 
unlike standard concatenation analysis under maximum likelihood they 
specifically address heterogeneity due to ILS.
Thus, they are in a distinctly different class of methods.

The method we mainly focus on here is SVDquartets
 \cite{svdquartets},   a linear algebra technique (based on the Singular Value Decomposition)  for computing a tree on a given set of four aligned 
sequences. 
SVDquartets has been recently proven to be a statistically consistent method for computing
quartet trees under
the MSC \cite{WascherKubatko-2019}.
Since SVDquartets only computes quartet trees, it must be combined with a quartet amalgamation method to estimate a tree on more than 4 species.
PAUP* \cite{paupstar} has one such approach, but another approach is
 SVDquest,  which uses a different technique to find good solutions to the optimization problem.

\paragraph{Computing quartet trees using SVDquartets}
Suppose the input is a multi-locus dataset, so that each locus is given as
an alignment of the four sequences, $a,b,c,d$.   SVDquartets computes statistics about the site patterns, and
then uses those statistics to determine which of the three trees is the best.
Hence, the output from SVDquartets on $a,b,c,d$ is one of the three possible
unrooted quartet trees $ab|cd$, $ac|bd$, or $ad|bc$.

\paragraph{Using SVDquartets on five or more species. } 
Now suppose the set $S$ of species for the input has more than four species.
To construct a tree on $S$, you do the following:
\begin{itemize}
\item[] Step 1: Use SVDQuartets to compute a tree on every four leaves, thus producing
a set $Q$ of quartet trees.
\item[] Step 2: Apply your preferred quartet amalgamation method to construct a tree on 
all of $S$ from $Q$.
\end{itemize}
Thus, the key to using SVDquartets on more than four species is the choice of
quartet amalgamation method.

PAUP* \cite{paupstar} contains an integrated approach to using SVDquartets
which uses a novel quartet amalgamation heuristic to combine the quartet tree.
The objective in the PAUP* heuristic is to find a species tree on the full set
$S$ of species that agrees with as many quartet trees in $Q$ as possible.
Since this is an NP-hard optimization problem \cite{jiang_polynomial-time_2001}, provably
optimal solutions are unlikely to be found using any heuristic, and 
(as with most phylogenetic estimation software) PAUP* uses 
local search strategies to find good solutions to this optimization problem.
Thus,  PAUP* can be used to compute trees under the SVDquartets
approach on any number of species.

However, any quartet amalgamation method can be used in Step 2,
and there are many possible quartet amalgamation methods to consider:
Quartets MaxCut \cite{snir_quartets_2008} and QFM \cite{reaz-quartet} are probably the
most well known, but see
also \cite{BryantSteel-quartets,piaggio2004quartet,quartet-joining-journal,quartet-joining-conf,ben1998constructing,holland-qimp,ranwez_quartet-based_2001} for some literature about
quartet amalgamation methods.
(Note that SVDquartets can also output the three possible quartets each with an associated statistic,
so that a weighted quartet amalgamation method, such as Weighted Quartets MaxCut \cite{avni2014weighted}, can be used).

\paragraph{SVDquest: improving the search strategy for SVDquartets.}
SVDquest \cite{svdquest} is a recent development for using SVDquartets on larger datasets that
uses a different kind of quartet amalgamation method than PAUP*.
Specifically, rather than employing a local search strategy to find a good
solution to the optimization problem, it uses the same {\em constrained
optimization} approach as used by ASTRAL, and then finds a provably
optimal solution within the constrained space.
Given the multi-locus set of
multiple sequence alignments, SVDquest takes the following steps:
\begin{itemize}
\item
Compute a tree on each locus (e.g., using maximum likelihood)
\item
Run ASTRAL on the set of trees to determine the set $X$ of allowed
bipartitions. 
\item
Run SVDquartets on the concatenated alignment for every four species  to compute the set $Q$ of quartet trees.
\item 
 Optionally, run PAUP* on $Q$ to obtain a tree $T^*$, and add its bipartitions to $X$.
\item
Find a tree $T$ on $S$ that agrees with the largest number of quartet trees  in $Q$, within the
constraint space defined by $X$ (i.e., $Bip(T) \subseteq X$ is required).
\end{itemize}
When the optional step is included, then by design
SVDquest is computationally more intensive
than PAUP* (since then it also calls PAUP*). In addition, it also computes
gene trees for every locus, which makes it more expensive again.
Thus, overall, it is substantially more expensive than PAUP*.

The question is therefore, why SVDquest would be used instead of PAUP* to
find a species tree? In essence there is only one reason: SVDquest is guaranteed to
find an optimal tree within its constraint space, which is the set of all binary trees whose
bipartitions are found in $X$.
Therefore, if SVDquest includes the optional step (i.e., runs PAUP* and adds the bipartitions from its tree into $X$),
SVDquest is guaranteed to produce a tree that agrees with {\em at least as many}
quartet trees as $T^*$, the tree found by PAUP*.
As shown in \cite{svdquest}, SVDquest often strictly improves
on PAUP* in terms of the criterion score, and so its use may be
valuable for some inputs.


The running time for SVDquartets to compute a quartet tree on a given set of
four species is linear in the number of sites in the alignment. 
Thus, for an alignment with   $n$ species and the length of the alignment is $M$, the
total running time to compute the set of quartet trees is $O(n^4 M)$.
Once the quartet trees are computed, then some quartet amalgamation heuristic  is
used. 
If PAUP* is used, then the running time is not predictable, since the heuristic
is based on local search strategies.
If SVDquest is used, the running time is essentially the same as 
the time to compute the quartet trees, compute the gene trees, and run an early version of ASTRAL.
Since this heuristic uses local search strategies, it's not really feasible to 
define the running time in any Big-O framework.
However, even just computing the set of all $O(n^4)$ quartet trees is
computationally intensive if $n$ is not sufficiently small. 

One approach to using SVDquartets on large datasets (where $n$ is 
sufficiently large that computing all possible quartet trees is too
expensive) is to sample quartets randomly, and compute quartet trees using
SVDquartets on only the sampled
quartets.
For example, you could select $n^2$ quartets randomly. 
It's possible that the random selection of quartets would miss some
species, in which case you could add in additional quartet trees to
ensure that all species are included in the dataset.
Finally, you would run the quartet amalgamation method on 
the set of quartet trees you obtained.
This approach has definite advantages in terms of running time, but
the accuracy of trees computed using sparsely sampled quartets may be reduced
compared to using the full set of quartet trees (see, for example, \cite{Swenson2011}).

\paragraph{Summary}
Overall, the site-based methods SVDquartets and SVDquest
(and most likely others)
are inherently well suited to very large numbers of loci.
The biggest challenge for these methods is when the number $n$ of species is large,
since the computation of $O(n^4)$ quartet trees quickly becomes infeasible.
Furthermore, as we discussed above,  sampling a smaller number of quartets to 
compute trees on makes the approach computationally feasible, but may reduce accuracy.
Therefore, site-based methods are a natural choice when the number of loci is very large and
the number of species is not more than about 100 (or perhaps even 200).

\section{Divide-and-conquer  for scaling species tree estimation methods}
\label{sec:divide-and-conquer}
Here we describe divide-and-conquer techniques for scaling species tree estimation
methods to large datasets.
Section \ref{sec:bbca} describes a very simple approach that can be used with *BEAST and other
co-estimation methods, when the number of loci is large.
Section \ref{sec:dtm} describes approaches that can be used when the
number of species is large, including very new approaches based on dividing the species into
disjoint subsets, constructing trees on subsets, and then merging the disjoint trees.
See \cite{warnow-festschrift-chapter} for additional discussion of these methods and issues.

\subsection{BBCA: scaling co-estimation methods to large numbers of loci}
\label{sec:bbca}

As shown in \cite{bbca}, the time needed for *BEAST to  reach good ESS values (used
as a prediction of whether the method had converged to the stationary distribution)
increased substantially with the number of loci, making analyses with 100 or more
loci computationally intensive, even for just 25 species. 
The 
BBCA \cite{bbca} technique is a very simple  approach for addressing this scalability challenge.
As noted in \cite{bbca,naive-binning}, the point estimates of gene
trees that *BEAST produces can be  more accurate than maximum likelihood gene trees.
Furthermore, species trees estimated using summary methods on
these *BEAST gene trees were as accurate as the species trees estimated using
*BEAST.
These observations suggested an approach for multi-locus species tree estimation
that is very simple, and that makes it possible to use *BEAST (and other co-estimation
methods) without the same computational effort.

\begin{itemize}
\item Step 1: Randomly partition the set of genes into ``bins" with the desired size (e.g., at
most 25 genes per bin).
\item Step 2: On each bin, use the desired co-estimation method (e.g., *BEAST) to produce estimated
gene trees for the genes in the bin.
\item Step 3: Run the desired summary method (e.g., ASTRAL) on the set of estimated
gene trees, to produce the estimated species tree.
\end{itemize}

Of these steps, the first is obviously very fast. The second step, in most cases,
will be the most expensive, since co-estimation methods such as *BEAST tend to be  computationally intensive, even on datasets with
only 25 genes. 
Most ways of running the third step will be generally fast (at least if a fast summary method is
used).
Therefore, the running time for a BBCA analysis is largely dominated by the second step.

BBCA was studied in \cite{bbca} on 
 simulated datasets with 11 species, 100 genes, and heterogeneity between true gene trees and species trees due to ILS.
True gene alignments, with varying sequence lengths, were randomly divided into 4 bins with 25 genes each. 
*BEAST was then run for 24 hours on each bin to produce gene trees, and
then the summary method MP-EST \cite{mpest} was used to combine these estimated gene trees into a species tree.
This BBCA analysis was  compared to *BEAST run on the full set of 100 genes for 96 hours (i.e., the same total amount of time);
 the result was that BBCA produced species trees that were at least as accurate as *BEAST species trees, but typically more accurate than *BEAST species trees.
 The explanation offered for the improved accuracy of BBCA over *BEAST
is that by only running *BEAST on 25 genes at a time, it was able to converge more quickly than on the full set of genes; this explanation is supported  by the improvement
in ESS values obtained for each of the bins compared to ESS values compared to ESS values for the full dataset.
In a second experiment, *BEAST was allowed to run for 168 hours (i.e., longer than the total BBCA time) on the full set of 100 genes. Even with the longer running time, *BEAST did not return a more accurate tree than BBCA.

The BBCA algorithm design has several parameters that can be adjusted by the user. 
For example, the algorithm is based on random partitioning into bins, but the choice of
the number of bins (or equivalently, the target number of genes in each bin) is up 
to the user.  Based on previous results,  
increasing the size of each bin most likely will improve accuracy but will also require more time for *BEAST to
converge.
Hence, the choice of the number of bins is a trade-off between running time and accuracy.
The first step of the BBCA algorithm is based on random partitioning of the loci into subsets, since the theoretical guarantee for *BEAST assumes that the loci it is given are drawn randomly from the genome; thus, random partitioning will provide this feature. 
On the other hand, it is certainly possible that empirical performance could be improved through non-random partitioning.  
In the second step of BBCA,  the user selects a co-estimation method to produce 
a distribution on gene trees for each locus in each bin; the BBCA algorithm was tested with *BEAST, but this could be
performed using StarBEAST2 or any other co-estimation method.
The user must also select the technique for producing the point estimate
of each gene tree given the gene tree distribution.
In the third step, a   summary method  is used to compute a species tree
from the set of gene trees computed in the second step.
BBCA was studied with MP-EST, but other summary methods could be used as well,
including ASTRAL, ASTRID, etc.

Thus, BBCA is a general framework for enabling computationally intensive co-estimation methods to run on datasets with large numbers of loci.
Furthermore, its algorithmic design, which is based on random partitioning  of the loci into small bins, 
means that it can be trivially parallelized, and since the second step
dominates the others in terms of running time, parallelization (by assigning
different bins to different processes) should result in a near-linear speed-up. 

An important limitation of BBCA is that it only produces a point estimate of the species
tree, rather than a distribution; hence, the use of 
BBCA instead of *BEAST does not provide the full power of a Bayesian method.
Nevertheless, when the dataset is too large (in terms of number of loci) to use 
*BEAST, then this technique can make it possible to have some of the
advantages of the co-estimation method without the computational hit.

\paragraph{Summary}
BBCA, has been used to scale *BEAST to large numbers of loci,
but it could be used with  any co-estimation method to run on
datasets with large numbers of loci.
BBCA thus addresses  
challenges imposed by large numbers of loci, but does not address the challenges 
when the number of species is large.

\subsection{Divide-and-conquer for analyzing large numbers of species}
\label{sec:dtm}

As we have discussed, most species tree estimation methods can be computationally 
intensive---and even ASTRAL can have difficulties completing analyses under
some conditions. 
Divide-and-conquer is an algorithmic approach 
that operates by 
dividing the species set into smaller subsets, constructing trees on the subsets, and then merging
the trees together into a tree on the full dataset.
These divide-and-conquer strategies can be used with ``base methods" to construct species trees
on subsets, thereby avoiding the challenges of running the species tree method on the
full set of species.
Divide-and-conquer  has been proposed as a potentially important
tool for large-scale species tree estimation in \cite{bininda-emonds-tree2004}, especially
for the case where the species set is divided into overlapping subsets,
trees are computed on the subsets, and then merged together using supertree methods
(see \cite{delsuc2005phylogenomics,bininda-emonds_phylogenetic_2004,sanderson2011terraces,fastrfs} for an entry into the literature
on supertree methods).
Here, we present a few divide-and-conquer approaches  that have been studied for use with
coalescent-based species tree estimation methods;
see
 \cite{warnow-festschrift-chapter} for additional discussion.
 

\paragraph{DACTAL. }
One of the earliest methods used for divide-and-conquer species tree estimation, DACTAL \cite{dactal},  was
originally studied in the context of computing trees from unaligned sequences.
In that context, it produces a tree on a set of unaligned sequences but does not
ever compute an alignment on the full set of sequences. 
As shown in \cite{dactal}, it produces more accurate trees than standard two-phase
methods (that first estimate an alignment and then compute an ML tree on the alignment)
on large datasets that are difficult to align.

DACTAL  was also used to improve species tree estimation in \cite{bayzid-dcm}, and so
we describe how it operates here.
DACTAL divides the dataset into overlapping subsets, constructs trees on subsets, and then 
merges these trees using  a supertree method (in \cite{dactal}, this was
performed using   SuperFine~\cite{SuperFine}, but other supertree methods could be used).
DACTAL combines iteration with divide-and-conquer, 
so that each iteration uses the
tree from the previous iteration to divide the species set into subsets.
As shown in \cite{bayzid-dcm}, using DACTAL with MP-EST to compute species trees on subsets
resulted in species trees that had higher accuracy than MP-EST species trees and reduced its running time. 

Although software for DACTAL is not available at this time,
the key idea is simple: divide the species set into overlapping subsets, construct
species trees on the subsets, and then combine the subset trees into a tree on the full set of
species using
a preferred supertree method; hence,
new divide-and-conquer strategies can be developed.
However, the design matters, as  the choice of  decomposition strategy  \cite{supertree-decomp} and the
supertree method \cite{nguyen2012mrl}  both have an impact on the accuracy of the final tree.

\paragraph{Disjoint Tree Merger techniques. }
More recently, a new divide-and-conquer strategy has been developed that
operates by dividing the dataset into {\em disjoint} subsets and then merges the
disjoint trees into a tree on the
full dataset.
Because the trees are disjoint, the step for combining disjoint trees requires extra information,
such as a distance matrix relating the species, and hence
cannot be accomplished by the use of a standard supertree method.
This step, which is referred to as {\em Disjoint Tree Merger} (DTM),
can be accomplished using various techniques,
including NJMerge \cite{njmerge-biorxiv}, TreeMerge \cite{molloy-ismb2019}, and constrained-INC \cite{ZhangRaoWarnow2018,Thien2019}.

NJMerge is a polynomial time modification of the neighbor joining method \cite{NJ} and  uses an input
distance matrix to 
agglomeratively build the tree through
siblinghood decisions, attempting at every point to only accept siblinghood
proposals that are consistent with the input constraint trees.
When NJMerge returns a tree, it tends to be highly
accurate, as shown in \cite{njmerge-biorxiv}.
However, because testing for compatibility of unrooted trees is NP-hard \cite{steel-compatibility}, 
NJMerge uses a heuristic to decide whether to accept a siblinghood proposal, and
on occasion it can accept a siblinghood proposal that leads to 
the subset trees being incompatible, in which case it will fail to return 
a tree.

TreeMerge  \cite{molloy-ismb2019} 
is another polynomial time DTM method that is an improvement on NJMerge in that it is guaranteed to always returns a tree that
satisfies the input constraint trees.
TreeMerge achieves this by running NJMerge only on pairs of constraint trees (where it never fails to correctly construct a tree), thus producing a new set of trees (on overlapping leaf sets) that satisfy the input constraint trees.
TreeMerge then estimates branch lengths for the trees produced by NJMerge so that they can be merged into a binary tree on the full leaf set that satisfies the input constraint trees.
TreeMerge thus has two advantages over NJMerge: it always outputs a tree that satisfies the constraint trees, and it is faster than NJMerge on large datasets
\cite{molloy-ismb2019}.
On the other hand, NJMerge produces slightly more accurate trees than TreeMerge on those datasets on which it completes.

NJMerge and TreeMerge have been evaluated in the context of species tree estimation. 
As seen in
\cite{njmerge-biorxiv,molloy-ismb2019}, both NJMerge and
TreeMerge tend to reduce the running time for all three species tree estimation
methods on large datasets (i.e., 1000 species and 1000 loci).
Furthermore, ASTRAL-III was unable to complete
many analyses on large datasets within 48 hours using 16 cores, but using NJMerge or TreeMerge 
enabled all analyses to complete using the same computational resources. 
For example, on high ILS datasets with 1000 species and 1000 genes,
 the entire divide-and-conquer pipeline using ASTRAL-III (to compute subset trees)
 and   TreeMerge (to combine subset trees)  completed in about 4 hours,
 whereas  using ASTRAL-III {\em de novo} (i.e., not
within the divide-and-conquer framework) typically did not
complete within 48 hours, and when it did complete it used close to the
48 hours allowed time on Blue Waters \cite{molloy-ismb2019}.
Finally, sometimes
accuracy was reduced when using the divide-and-conquer approach,
but when this happened the reduction in accuracy was very small.

The other DTM method  that we mentioned is constrained-INC.
However, unlike
NJMerge and TreeMerge, constrained-INC has only been studied in the context of gene tree
estimation under the 
GTR (Generalized Time Reversible) model \cite{tavare-gtr} of sequence
evolution; hence, it is not known how well constrained-INC will perform when
used with species tree estimation methods.

\section{Parallel implementations for phylogenomic analyses}
\label{sec:parallel}

We have discussed scalability in the context of asymptotic running time, which does not take into account whether some operations can be performed simultaneously.
Indeed, a different perspective on scalability examines the impact of using additional processors; for example, can a species tree estimation method run in less time if given access to a greater number of processors?
Ideally, running a method using $p$ processors (instead of one processor) would decrease the running time by a factor of $p$, corresponding to 100\% parallel efficiency.
However, in practice, some operations depend on the previous operations having been performed (i.e., the work is serial), and in this case, parallel efficiency will be less than 100\%.
Amdahl's law \cite{amdahl1967validity} states that parallel efficiency is governed by the fraction of serial work, and in particular, when the fraction of serial work is greater than 0, parallel efficiency goes to 0 as the number of processors goes to infinity.
In other words, as the number of processors increases, the execution time becomes dominated by the serial work!
Although disappointing, this observation is useful for evaluating the scalability of parallel algorithms; see review in \cite{heath2015a}.
Here, we discuss recent advances in parallel codes for large-scale phylogenomic analyses with an emphasis on the serial work performed by these methods to illustrate open challenges.

\subsection{Maximum Likelihood on multi-locus datasets}

Maximum likelihood analysis is statistically consistent when all the sites evolve
down a common GTR model tree, but  different loci in a multi-locus dataset are expected to have
different tree topologies due to ILS, and so the statistical consistency guarantee no longer
holds (and in fact, concatenation analyses using maximum likelihood can be positively
misleading \cite{RochNuteWarnow2018}). However, maximum likelihood is a common approach  to multi-locus species tree estimation, and so we discuss the computational issues for this kind of analysis in this section.

\paragraph{RAxML. } 
RAxML is one of the most widely used methods for phylogenomic analyses, and parallel versions of RAxML have existed for many years.
Recall that ML heuristics take as input a multiple sequence alignment and perform a heuristic search of tree space, computing the likelihood for candidate trees.
Because the likelihood of observing a site pattern given a candidate tree is {\em independent} of the other sites in the alignment, the log-likelihood for each site can be computed in parallel and then summed together.
Parallelism across sites can be implemented at many different levels, and RAxML version 8 \cite{raxml8} uses pthreads, vector extensions (SSE3, AVX and AVX2), and other techniques to reduce the amount of time required to compute tree log-likelihood, which can be computationally intensive when there are many unique site patterns in an alignment.
Importantly, these optimizations do not impact the tree search (although note that {\bf RAxML-NG} \cite{kozlov2019raxmlng}, a recently released version of RAxML, re-implements the tree search algorithm in order to improve speed and provide additional functionality).
RAxML version 8 can search from multiple starting trees in parallel (using MPI); however, this parallelism does not reduce the number of candidate trees that need to be evaluated for any one of the searches to converge to a local optimum, and searching tree space is effectively serial work.
Because treespace increases exponentially with the number of species,  the search phase may be longer for datasets with large numbers of species.
Thus, despite significant optimizations, running RAxML can be computationally intensive for some datasets.

\paragraph{ExaML. } For some very large multi-locus datasets, the concatenated alignment may not fit into the memory of a single compute node.
In this case, researchers with access to a distributed-memory system can run a different MPI version of RAxML, called ExaML \cite{kozlov2015examl}.
ExaML operates by partitioning the alignment across sites and distributing these partitions across multiple compute nodes.
Thus, computing tree log-likelihood for the {\em entire} alignment and coordinating the tree search requires some amount of communication (i.e., the sending/receiving of messages) between processes.
For example, it would require $\log_2(p)$ steps to sum the log-likelihood across all alignment partitions using a standard global reduction on $p$ processes, and these $\log_2(p)$ steps are effectively serial work.
Furthermore, the amount of time required for communication is significant compared to other operations, which is why   avoiding communication \cite{demmel2013communication}, overlapping communication and computation  \cite{hoefler2007implementation}, and modeling communication  \cite{culler1993logp, alexandrov1995loggp, gropp2016modeling} are topics of interest to the high performance computing community.
Finally, ExaML faces the same challenges as RAxML when it comes to effectively searching tree space, and thus, although ExaML is a significant advancement in large-scale ML tree estimation, there are still open challenges.

\paragraph{Concatenation versus gene tree estimation followed by summary methods. }

The major competing approaches in multi-locus species tree estimation, at this time, are
concatenation analyses using maximum likelihood or gene tree estimation followed by a summary method.
Because  many summary methods are very fast (e.g., ASTRAL and ASTRID), from a computational viewpoint, the difference between concatenation or summary methods on estimated gene trees comes down to  whether it is faster or slower to compute $k$ gene trees rather than a single tree on the concatenated multiple sequence alignment for the $k$ loci.

For simplicity, we assume that the tree search is the same regardless of the approach, that is, the same candidate trees are evaluated in the same order for the concatenated alignment as well as each of the multiple sequence alignments for the $k$ different loci.
We also suppose that the concatenated alignment is large enough that ExaML must be used to perform the analysis, but the alignment of each locus is small enough that RAxML can be used to perform each gene tree analysis, which was the case in the Avian Phylogenomics Project \cite{jarvis-2014b}.
Thus, each gene tree analysis (using RAxML) can performed in an embarrassingly parallel fashion, but the concatenation analysis (using ExaML) will require some amount of communication to coordinate the tree search across multiple processes.
Because the cost of communication is significantly greater than the cost of other operations (e.g., floating-point arithmetic), we predict that the concatenation analysis will take longer than analyzing each of the gene trees independently using the same computational resources---as these analyses can be performed in parallel, but without communication.
It is also possible that analyzing each loci individually could result in better memory locality (e.g., cache-oblivious algorithms \cite{frigo1999cache}), further improving performance.
Finally, as different parts of the genome can support different model trees, the tree search may converge to a reasonable local optimum more slowly for the concatenated alignment compared to the alignments for individual loci.

This analysis shows that there are several reasons that one would expect it to be faster to compute $k$ gene trees compared to a single tree on the concatenated alignment for the $k$ loci.
Furthermore, improvements in ML methods for gene tree estimation, including the recently introduced method ParGenes \cite{kozlov2018pargenes}, are likely to tip the balance further in favor of gene tree estimation followed by summary methods. 
While concatenation analyses often provide excellent accuracy and are still the norm for biological systematics, for those datasets with thousands (or tens of thousands) of loci,  gene tree estimation followed by summary methods are likely to be computationally more efficient and  provide the same level of accuracy under many conditions (and are sometimes more accurate).

\subsection{ASTRAL-MP}

ASTRAL-MP is a   recent development within the ASTRAL suite of methods  \cite{astral-mp}.
While previous versions of ASTRAL used only a single thread, 
ASTRAL-MP implements parallelism using  vectorization (AVX2),
CPU multi-threading, and GPU multi-threading (OpenCL). 
To achieve high scaling with large numbers of cores, ASTRL-MP changes some parts of the ASTRAL algorithm, but does this without sacrificing  statistical consistency.
AVX2 vectorization and CPU multi-threading enable speed-ups (compared to ASTRAL-III) on most modern machines, and users with access to GPUs machines will benefit from even greater speed-ups. For example, the GPU version of ASTRAL-MP achieved speed-ups of 158X (compared to ASTRAL-III) and enabled the analysis of a dataset with 147,800 gene trees and 144 species in less than two days; in contrast, an approximation of the ASTRAL-III analysis for this dataset used 180 days~\cite{astral-mp}.
Datasets with larger numbers of species (10,000 species and 1,000 genes) were also able to be analyzed in less than two days.
 In summary, GPU version of ASTRAL-MP enables fast species tree estimation on datasets with large numbers of species (but not too many genes) or large numbers of genes (but only a moderate number of species), and is an exciting development in large-scale species tree estimation.

\section{Choosing between methods}
\label{sec:discussion}
This chapter  explored different methods for estimating species trees given
multi-locus datasets where true gene trees can differ from the true species tree (and
from each other) due to incomplete lineage sorting.
Our focus has been on their accuracy and computational requirements when constructing species trees for large datasets,
where datasets can be large  in the number $n$ of species or the number $k$ of loci (or both!).


When  the number $n$ of species is sufficiently small, then nearly all the methods we discussed (i.e., ASTRAL, ASTRID, SVDquartets, SVDquest, and concatenation using maximum likelihood) can run, although 
*BEAST will have problems unless the number of loci is very small. 
BBCA is a divide-and-conquer technique that
enables *BEAST (and other methods for co-estimating gene trees and
species trees) to scale to datasets when $k$ is large, but does
not address the computational challenge when $n$ is large.
Moreover, even when $n$ is relatively small (e.g., under $100$), 
concatenation analyses using the better maximum
likelihood codes, such as RAxML or ExaML, can be computationally intensive, requiring
terabytes of shared memory, years of CPU time,  and the use of supercomputers (as
our experience with the Avian Phylogenomics Project analyses revealed).
In contrast, the summary methods  (ASTRAL and ASTRID) and site-based methods (SVDquartets within PAUP*
and SVDquest) remain
practical techniques for  analyzing these datasets.
The relative accuracy of these methods is still being explored,
but all of these methods are fast enough to be used on datasets of this size.
A natural approach to species tree estimation for such datasets would be to
run all these  analyses (e.g., ASTRID, ASTRAL, SVDquartets or SVDquest, and
concatenation analyses) and then  
examine the resulting trees for common features.
Furthermore, since concatenation analyses will tend to be the most computationally
intensive, the other methods could be run first, and then concatenation analyses could
be used if necessary, based on whether the evolutionary questions of interest
are answered adequately using these analyses.

As $n$ increases, the set of methods that are able to complete using standard
resources (i.e., without supercomputers) decreases, and in particular
concatenation analyses using maximum likelihood can become infeasible
without substantial time on supercomputing platforms.
The current research, described here, suggests that in these cases, 
the better summary methods (notably ASTRAL, and possibly ASTRID) can provide good accuracy
and can be fast enough (with low enough memory requirements) to complete within reasonable times and without
supercomputers for many datasets.
The main effort in analyses using ASTRAL or ASTRID (or other summary methods) is the calculation
of gene trees, which depends also on the calculation of multiple sequence alignments;
these are both computationally and statistically challenging problems, but methods such
as PASTA \cite{Mirarab2014}  (which co-estimates alignments and trees) can be highly accurate and fast, even on datasets with several thousand
sequences.
Assembling the species tree from the estimated gene trees is then generally very fast, using in some cases a few CPU days of analysis, but much less than the CPU years used by concatenation analyses.

For very large $n$, however, nearly all methods become either
infeasible to use without extensive resources or require substantial modification.
For example, concatenation analysis using ML (if performed using RAxML or similarly
accurate but highly computationally intensive methods) becomes infeasible to run to
completion without extensive computational resources (such as substantial memory), 
methods based on SVDquartets need to be modified to not compute all quartet trees,
and
 even ASTRAL may
not complete within reasonable timeframes.
As we noted in Section \ref{sec:parallel}, parallelism can speed up analyses but
does not address the challenges of large numbers of species, as treespace
increases exponentially with the number $n$ of species.
This is the context in which the divide-and-conquer strategies we discussed are likely to be
very useful. Furthermore,  the Disjoint Tree Merger (DTM) methods NJMerge and TreeMerge have both been studied in the
context of multi-locus species tree estimation and were able to improve scalability of species tree
estimation methods, including ASTRAL-III and 
concatenation using maximum likelihood,
and produced
trees with high accuracy in reduced time.

\section{Future Directions}
\label{sec:conclusions}
Phylogeny estimation is well known to be one of the most computationally challenging 
problems in the biological sciences, as most of the problems are NP-hard, and 
this is also true for  optimization problems posed for species tree estimation.
Furthermore, because 
of gene tree heterogeneity and the benefit of dense taxon sampling strategies, 
multi-locus datasets are often large in both numbers of loci and also number of species, making
for particularly challenging analyses.

In this chapter, we have compared some of the leading methods for species tree estimation in the presence of ILS with
respect to accuracy and running time on large and ultra-large datasets,
mainly focusing on the challenges that ensue when the number of species is very large.
We showed that summary methods, such as ASTRAL and ASTRID, provide improved
scalability compared to concatenation analyses using maximum likelihood, and can
provide high accuracy when gene trees have good accuracy. 
In contrast, when gene trees are poorly resolved, although very computationally
expensive on large datasets, concatenation analyses and
site-based methods, such as SVDquartets, can provide advantages. 
Divide-and-conquer strategies based on Disjoint Tree Merger (DTM) techniques, such as TreeMerge, can enable concatenation with maximum likelihood  and site-based methods to scale to large
datasets, with minimal impact on accuracy.
Thus, highly accurate ultra-large species tree estimation may be feasible using a combination of strategies.

However, the development of DTM methods are in their very early stages, and most
likely there will be substantial improvements in their algorithmic design that impact
accuracy, running time, and scalability to large datasets.
Similarly, although no functional implementation currently exists for DACTAL, divide-and-conquer 
strategies that rely on supertree methods (rather than DTM methods) are also
likely to be valuable tools for scaling species tree estimation methods
to large datasets.
Thus, biologists who are considering assembling ultra-large datasets may
well benefit from future algorithmic developments, and are encouraged to keep an eye out for new software that enable species tree methods to scale to large datasets.

This chapter suggests several directions for future work. 
For example, we did not discuss how to estimate branch support for ultra-large species trees,
A relatively simple and computationally feasible approach that can be used when gene trees are available is to give the estimated species tree and the estimated gene trees to ASTRAL, and let it compute the posterior probabilities on the branches of the estimated species tree.
Alternative approaches, perhaps based on bootstrapping, can also be used but will be more
expensive and may not provide the same level of accuracy \cite{MirarabSysBio2014}.
The incidence of missing data (i.e., gene trees that lack species) is likely to increase on large species datasets, and while some studies have suggested that the leading species tree methods are 
robust to many patterns of missing data \cite{Molloy2018,Nute-missing2018}, 
this question has not been carefully examined on datasets with large numbers of species.
Gene tree estimation error is also known to impact species tree estimation when performed
using summary methods \cite{Molloy2018}, and efforts to improve
gene trees in a multi-locus setting should be developed.
Statistical binning \cite{statbinning} and its improvement, 
weighted statistical binning \cite{WSB}, can be seen as attempts to improve
gene tree estimation (and thereby species tree estimation), but direct
co-estimation using *BEAST and similar techniques is likely to 
provide better results; however, as noted, co-estimation 
methods  that can scale to large and ultra-large 
numbers of species have not yet been developed.

Finally, this chapter specifically focused on multi-locus species tree estimation when gene trees
can differ from each other and from the true species tree due to ILS.
Other biological processes, such as gene duplication and loss,  also result in gene tree 
discordance and
complicate species tree estimation.
Species tree estimation in the presence of gene duplication and loss can be computationally
intensive (e.g., Phyldog \cite{Boussau-genome} is very expensive), and could potentially be scaled
to large datasets when used within
the divide-and-conquer strategies we discussed here. 
Events such as hybrid speciation and horizontal gene
transfer require methods for phylogenetic network estimation, which are much
more computationally intensive than the phylogenomic estimation methods
we have described here.
However,  the divide-and-conquer
approaches we presented are not directly relevant to phylogenetic network estimation,  since by construction they assume that the input is a set of leaf-disjoint trees. 
However,   divide-and-conquer approaches have been  developed (surveyed in \cite{Zhu2019}) for scaling phylogenetic network
estimation methods to large datasets, and 
one of these has been
integrated into Phylonet \cite{phylonet-bmc2008}, a software
suite of different methods for phylogenetic network estimation.

\section*{Acknowledgments}
This work was supported in part by NSF grant  CCF-1535977.

\appendix
\section{Big-O analysis}
An introduction to {\em Big-O running time analysis} can be found in many places
(e.g.,  in Wikipedia, and also in
 \cite{warnow-textbook}),  but we provide a brief example here.
Let $f$ and $g$ be two real-valued functions (i.e., $f$ and $g$ both map the real
numbers to real numbers), and suppose for the sake of simplicity that both 
map positive real numbers to positive real numbers. 
Note that $f$ is the name of the function and $f(n)$ is the value of the function on input value $n$.

We say that $f$ is ``Big-O" of $g$ (written ``$f$ is $O(g)$") if
there is a pair of constants $C, C'$ such that $f(n) \leq Cg(n)$ for all $n \geq C'$.
In other words, when $n$ is large enough, then $f(n) $ is bounded from above by
$C g(n)$.
Thus, Big-O ignores multiplicative or additive constants, and provides an upper bound on the 
growth of a function.

Some examples may help clarify this definition.
For example, if $f(n)=5n+1000$ and $g(n)=n^2$, then $f(n) \leq 6g(n)$ if $n \geq 1000$, and
so $f$ is $O(g)$.
But we would also say that $f$ is $O(h)$,  where $h(n) = n$, since $f(n) \leq 6h(n)$ for large
enough $n$.
We can also express this by saying $f$ is $O(n)$, which is easier to parse.

Now consider two functions  $f(n)=5n$ and $g(n)=n-3$. 
Is it the case that   $f$ is $O(g)$? The answer is yes, since $f(n) \leq 6g(n)$ for $n$ large enough. Note therefore that $f$ is $O(g)$ in this case, even though $f(n)$ is always strictly bigger than $g(n)$.
Also, $n^2+1000n$ is $O(n^2)$ but it is also $O(n^3)$ and $O(n^4)$.
Thus, 
saying that a function $f$ is $O(g)$  only expresses that $f(n)$ is asymptotically 
{\em bounded from above} by some constant times $g(n)$, and the key points are that it is
asymptotic (so only depends on being true for large enough values of $n$) and that it's an upper bound.

Hence, when we talk about running time analysis using Big-O notation, we try to use the tightest and simplest upper bound we can.
Thus, we will prefer to say that $f(n)=3n^2+1000n$ is $O(n^2)$ rather than $O(n^3)$ (although 
both statements are true), because $O(n^2)$ is a tighter upper bound than $O(n^3)$.
Similarly 
we will not want  to say $f(n)$ is $O(3n^2)$ (even though that's true) because
it's not as simple as saying $f(n)$ is $O(n^2)$.
Thus, we aim for a tight upper bound that is simple to express, noting again that the definition of 
Big-O doesn't care about multiplicative or additive constants.

This definition of ``Big-O" let's us discuss running time of different methods, by
expressing the running time as a function $f(n)$ where $n$ is the size of the input.
For example, suppose that we have two different methods that both take an array of $n$ integers as input.
We will consider every operation (i.e., numerical operations, logical operations, reads, 
writes, etc.) as having the same cost (yes, this is not really true, but it simplifies the
analysis, and is how running times are generally analyzed).
Then, if method A performs $n^2 + 3n + 10000$ operations, we can say that method A has $O(n^2)$ running time (because the most important part of the running time is the $n^2$ part); similarly, if method B performs $1000n$ operations, then method B has $O(n)$ running time.
From a ``Big-O" perspective,   it is easy to see that method B is more scalable than method A, since for large enough values of $n$ the running time for $B$ will be lower than the running time for $A$ (i.e., just think about the difference between $1000n$ and $n^2$ when $n=1,000,000$)!
Importantly, asymptotic running time does not depend on experimental details (e.g., the language used to implement a method or the computer system used to run the analyses), and enables a comparison of methods based on the size of the input.

\section{Software}

We provide information about the software discussed in this paper.

\paragraph{ASTRAL}
\begin{itemize}
\item
Software available in open-source form at \url{https://github.com/smirarab/ASTRAL}
\item
Tutorial at \url{https://github.com/smirarab/ASTRAL/blob/master/astral-tutorial.md}
\item
Main publications: \cite{astral,astral2,astral3,astral-mp}
\item
Comments: The main options that impact ASTRAL's accuracy and running time are those that impact the definition of the constraint set $X$ (larger sets can improve accuracy but also increase the running time) and whether low support branches in the gene trees are collapsed.
The most recent version, ASTRAL-MP uses a parallel compute cluster to enable ASTRAL to 
run on much larger datasets. 
 The largest datasets that ASTRAL analyzed in \cite{astral2} have 1000 species
and 1000 estimated gene trees.
In \cite{astral3}, ASTRAL was studied on much larger datasets with up to
10,000 species and 1000 gene trees, and most analyses successfully completed.
\item
Contact: astral-users@googlegroups.com  
\end{itemize}

\paragraph{ASTRID}
\begin{itemize}
\item Software available in open-source form at \url{https://github.com/pranjalv123/ASTRID}
\item Main publication: \cite{astrid} 
\item Comments: Although the calculation of the internode distance matrix is specified,
the calculation of the tree from the distance matrix can be performed using
several different options.
In the original study, FastME was used to construct trees and had good accuracy.
Other distance-based methods can also be employed, and may provide
improved accuracy.
The largest datasets that ASTRID analyzed in \cite{astrid} have 1000 species
and 1000 estimated gene trees. 
\item Google users group: \url{https://groups.google.com/forum/#!forum/astrid-users}
\end{itemize}

\paragraph{BBCA}
\begin{itemize}
\item
Software: There is no software for BBCA, because the approach is simple enough that no
software is necessary.
\item 
Comments: The BBCA algorithmic design leaves three important decisions up to the
user: the number of loci per bin, the co-estimation method used on
each bin, and the preferred
 summary method.
 To maximize accuracy reducing the number of bins is most likely best, but the time needed
 to converge on each bin will increase.
 In addition, although not explored, 
instead of assigning the loci randomly to bins (which maintains heterogeneity
within bins),
loci could be assigned to bins to reduce heterogeneity within bins.
The largest datasets analyzed by BBCA in \cite{bbca} have 100 genes and 12 loci.
\end{itemize}

\paragraph{Constrained-INC}
\begin{itemize}
\item Software available at
\url{https://github.com/steven-le-thien/constraint_inc}
\item Main publications: \cite{Thien2019}
\item Comments: Constrained-INC is designed to combine disjoint trees, so
needs to be used in a divide-and-conquer protocol to enable species tree estimation.
As a result, the main question is how the species dataset is divided into disjoint
subsets and how species trees are constructed on subsets.
Constrained-INC has only been studied in the context of GTR gene tree
estimation.
The largest datasets analyzed by Constrained-INC have 10,000
sequences. 
\end{itemize}

\paragraph{NJMerge}
\begin{itemize}
\item
Software available in open-source form at \url{https://github.com/ekmolloy/njmerge}
\item 
Main publications: \cite{njmerge,njmerge-biorxiv}
\item
Comments:  NJMerge is designed to combine disjoint trees, and so 
needs to be used within a divide-and-conquer strategy to construct species trees.
As a result, the main question is how the species dataset is divided into disjoint
subsets and how species trees are constructed on subsets.
 NJMerge was used with ASTRAL, RAxML, and
SVDquartets in \cite{njmerge-biorxiv}; the largest of these datasets have
1000 species and 1000 loci. 
\end{itemize}

\paragraph{StarBEAST2}
\begin{itemize}
\item Software available at \url{https://github.com/genomescale/starbeast2}.
\item Tutorial at \url{https://taming-the-beast.org/tutorials/starbeast2-tutorial}
\item Main publication: \cite{starbeast2}

\end{itemize}

\paragraph{SVDquartets within PAUP*}
\begin{itemize}
\item Software for PAUP* available at \url{http://paup.phylosolutions.com}
\item Tutorial at: \url{http://www.phylosolutions.com/tutorials/ssb2018/svdquartets-tutorial.html}
\end{itemize}

\paragraph{SVDquest}
\begin{itemize}
\item Software available in open-source form at \url{https://github.com/pranjalv123/SVDquest}
\item Main publication: \cite{svdquest}
\item Comments: SVDquest works with ASTRAL to help define the constraint set $X$, and
its running time and accuracy is similarly impacted by how $X$ is defined.
One important consideration is whether the bipartitions from the SVDquartets tree produced by PAUP* are included in the set $X$; 
including these bipartitions ensures that the optimization criterion score produced by SVDquest is at least as good as the score produced by PAUP*, but adds to the running time.
  The study introducing SVDquest \cite{svdquest} explored  simulated datasets with up to 50 species
and up to 1000 loci (with 300-1500 sites per locus) and a 37-species mammalian dataset with
424 loci and a total of 1,338,678 sites.
\end{itemize}

\paragraph{TreeMerge}
\begin{itemize}
\item
Software available in open-source form at \url{https://github.com/ekmolloy/treemerge}
\item
Main publication: \cite{molloy-ismb2019}
\item
Comments:  TreeMerge is used within a divide-and-conquer strategy.
As a result, the main question is how the species dataset is divided into disjoint
subsets and how species trees are constructed on subsets.
TreeMerge was used with ASTRAL and concatenation using maximum likelihood (computed
using RAxML) in \cite{molloy-ismb2019}; the largest of these datasets have
1000 species and 1000 loci. 
\end{itemize}

\bibliographystyle{plain}
\bibliography{kk-v2}

\end{document}